\documentclass[
reprint,
superscriptaddress,
bibnotes,
amsmath,amssymb,
aps,
prb,
longbibliography
]{revtex4-2}

\usepackage{graphicx,import}
\usepackage{dcolumn}
\usepackage{braket}
\usepackage{textcomp, gensymb}
\usepackage{bm}
\usepackage{amsmath}
\usepackage{float}
\usepackage{mathtools}
\usepackage{color}
\usepackage[dvipsnames,table,xcdraw]{xcolor}
\usepackage[utf8]{inputenc}
\usepackage{csquotes}
\usepackage{thmtools}
\usepackage{here} 
\usepackage{dsfont}
\usepackage{bbm}

\usepackage{hyperref}

\usepackage[normalem]{ulem}
\usepackage[left]{lineno} 
\usepackage{outline}
\usepackage{physics}
\usepackage{multirow}
\usepackage{siunitx}
\usepackage{tikz}
\usetikzlibrary{shapes.geometric, arrows,positioning}

\usepackage{qcircuit}
\usepackage{lipsum}
\usepackage{chemfig}
\usepackage{silence}
\usepackage{algorithm}
\usepackage{algpseudocode}
\usepackage{upgreek}

\makeatletter
\def\p@subsection{}
\makeatother
\makeatletter
\def\p@subsubsection{}
\makeatother

\allowdisplaybreaks

\begin{document}

\title{Concatenated continuous driving of silicon qubit by amplitude and phase modulation}

\author{Takuma Kuno}
\email{takuma.kuno.pg@hitachi.com}
\affiliation{Research and Development Group, Hitachi, Ltd., Kokubunji, Tokyo 185-8601, Japan}
\affiliation{Department of Electrical and Electronic Engineering, Institute of Science Tokyo, Meguro, Tokyo
  152-8552, Japan}

\author{Takeru Utsugi}
\affiliation{Research and Development Group, Hitachi, Ltd., Kokubunji, Tokyo 185-8601, Japan}

\author{Andrew J. Ramsay}
\affiliation{Hitachi Cambridge Laboratory, J. J. Thomson Ave., Cambridge, CB3
  0HE, United Kingdom}

\author{Normann Mertig}
\affiliation{Hitachi Cambridge Laboratory, J. J. Thomson Ave., Cambridge, CB3
  0HE, United Kingdom}

\author{Noriyuki Lee}
\affiliation{Research and Development Group, Hitachi, Ltd., Kokubunji, Tokyo 185-8601, Japan}

\author{Itaru Yanagi}
\affiliation{Research and Development Group, Hitachi, Ltd., Kokubunji, Tokyo 185-8601, Japan}

\author{Toshiyuki Mine}
\affiliation{Research and Development Group, Hitachi, Ltd., Kokubunji, Tokyo 185-8601, Japan}

\author{Nobuhiro Kusuno}
\affiliation{Research and Development Group, Hitachi, Ltd., Kokubunji, Tokyo 185-8601, Japan}

\author{Hideo Arimoto}
\affiliation{Research and Development Group, Hitachi, Ltd., Kokubunji, Tokyo 185-8601, Japan}

\author{Sofie Beyne}
\affiliation{IMEC, 3001 Leuven, Belgium}
\author{Julien Jussot}
\affiliation{IMEC, 3001 Leuven, Belgium}
\author{Stefan Kubicek}
\affiliation{IMEC, 3001 Leuven, Belgium}
\author{Yann Canvel}
\affiliation{IMEC, 3001 Leuven, Belgium}
\author{Clement Godfrin}
\affiliation{IMEC, 3001 Leuven, Belgium}
\author{Bart Raes}
\affiliation{IMEC, 3001 Leuven, Belgium}
\author{Yosuke Shimura}
\affiliation{IMEC, 3001 Leuven, Belgium}

\author{Roger Loo}
\affiliation{IMEC, 3001 Leuven, Belgium}
\affiliation{Department of Solid-State Sciences, Ghent University, Krijgslaan 285, 9000 Ghent, Belgium}

\author{Sylvain Baudot}
\affiliation{IMEC, 3001 Leuven, Belgium}
\author{Danny Wan}
\affiliation{IMEC, 3001 Leuven, Belgium}

\author{Kristiaan De Greve}
\affiliation{IMEC, 3001 Leuven, Belgium}
\affiliation{Department of Electrical Engineering (ESAT-MNS), KU Leuven, Leuven, Belgium}

\author{Shinichi Saito}
\affiliation{Research and Development Group, Hitachi, Ltd., Kokubunji, Tokyo 185-8601, Japan}

\author{Digh Hisamoto}
\affiliation{Research and Development Group, Hitachi, Ltd., Kokubunji, Tokyo 185-8601, Japan}

\author{Ryuta Tsuchiya}
\affiliation{Research and Development Group, Hitachi, Ltd., Kokubunji, Tokyo 185-8601, Japan}

\author{Tetsuo Kodera}
\affiliation{Department of Electrical and Electronic Engineering, Institute of Science Tokyo, Meguro, Tokyo
  152-8552, Japan}

\author{Hiroyuki Mizuno}
\affiliation{Research and Development Group, Hitachi, Ltd., Kokubunji, Tokyo 185-8601, Japan}

\date{\today}

\begin{abstract}

  The rate of coherence loss is lower for a qubit under the Rabi drive than a freely evolving qubit, $T_{2}^{\rm{Rabi}}>T_{2}^*$.
  Building on this principle, concatenated continuous driving (CCD) keeps the qubit under continuous drive to suppress noise and manipulate dressed states by either phase or amplitude modulation.
  In this work, we propose a variant of CCD which simultaneously modulates both the amplitude and phase of the driving field to generate a circularly-polarized field in the rotating frame of the carrier frequency.
  This circular-modulated CCD (CMCCD) cancels the counterrotating term in the second rotating frame, eliminating a systematic pulse-area error that arises from an imperfect rotating wave approximation for fast gates.
  Numerical simulations demonstrate that the proposed CMCCD achieves higher gate fidelity than conventional CCD schemes.
  We further implement and compare different CCD protocols using an electron spin-qubit in an isotopically purified $^{28}$Si-MOS quantum dot and evaluate its robustness by applying static detuning and Rabi frequency errors.
  The robustness is significantly improved compared with the standard Rabi drive, showing the effectiveness of this scheme for qubit arrays with variation in qubit frequency, coupling to the Rabi drive, and low-frequency noise.
  The proposed scheme can be applied to various physical systems, including trapped atoms, cold atoms, superconducting qubits, and NV centers.

\end{abstract}

\maketitle

\section{Introduction}
\label{sec:introduction}

The long coherence time of a quantum state is an essential resource for quantum technologies, including quantum sensing and metrology~\cite{degen2017quantum,demille2024quantum,giovannetti2006quantum}, quantum simulations~\cite{feynman1982simulating,georgescu2014quantum,hensgens2017quantum}, and quantum computing~\cite{divincenzo2000physical,nielsen2010quantum}.
Quantum computation requires the high-precision control of large-scale qubit arrays~\cite{raussendorf2007fault,fowler2012surface,preskill2018quantum}.
Low-frequency noise can be considered an uncertainty in the qubit frequency and drive field.
Consequently, control methods that are tolerant to uncertainties are important.
Furthermore, the inhomogeneity of the drive field and the variations of the qubit characteristics across the large-scale qubits pose significant challenges to achieving reliable qubit control.

Dressing of a two-level quantum state has been demonstrated to effectively extend coherence times in various quantum systems, including trapped atoms~\cite{timoney2011quantum,tan2013demonstration}, NV centers~\cite{xu2012coherence,golter2014protecting,morishita2019extension}, silicon carbide divacancy~\cite{miao2020universal}, donor-based qubit in silicon ~\cite{laucht2017dressed}, and hole-qubit in germanium~\cite{,tsoukalas2025dressed}.
In the dressed state, an electromagnetic driving field interacts resonantly with the quantum system and induces dynamical decoupling from noise, providing reliable control of the quantum states.
Although this approach has the advantage of being robust against environmental noise, it remains vulnerable to fluctuations in the driving field.
One approach to address this issue is the sinusoidally modulated, always rotating and tailored (SMART) protocol, proposed in Refs.~\cite{hansen2021pulse,seedhouse2021quantum}.
The SMART protocol utilizes the alternating phase of a sinusoidally amplitude-modulated drive to reduce the impact of drive field fluctuations and detuning error.
The SMART protocol has been demonstrated using qubits in NV centers~\cite{vallabhapurapu2023high} and silicon quantum dots~\cite{hansen2022implementation} and has recently been effective for the entangling gate operation~\cite{hansen2024entangling}.
Another strategy is the concatenated continuous driving (CCD) scheme~\cite{cai2012robust,cohen2017continuous,stark2017narrow,farfurnik2017experimental,wang2020coherence,cao2020protecting,wang2021observation,salhov2024protecting}.
By modulating the microwave (MW) driving amplitude or phase, the CCD scheme engages the quantum states in the double-dressed state, decreasing sensitivity to the driving field fluctuations.
The CCD scheme has recently been experimentally studied for the qubit application~\cite{ramsay2023coherence} and demonstrated gate fidelity improvement in noisy isotopically natural Si-MOS devices~\cite{kuno2026robust}.
Since the double-dressed qubit requires control in the second rotating frame, the rotating wave approximation (RWA) is needed to cancel the counterrotating term.
However, the residual counterrotating term can be an obstacle for precise qubit control, resulting in decreased gate fidelity.
The validity of the RWA can also become important in these dressed-qubit-based approaches, particularly under strong-driving or fast-control conditions~\cite{seedhouse2021quantum,hansen2022implementation,vallabhapurapu2023high}.

In this work, we propose a variant of the CCD protocol.
Unlike conventional methods, which employ either amplitude or phase modulation, our approach modulates both
amplitude and phase simultaneously.
This circular-modulated CCD (CMCCD) cancels the counterrotating field, resulting in a circular polarization driving field in the first rotating frame (Fig.~\ref{C0}).
Consequently, in the second rotating frame of the modulated drive, there is no residual second harmonic term to give rise to a systematic gate error on resonance.
We experimentally demonstrate the proposed CMCCD using a spin-qubit in an isotopically enriched silicon quantum dot.
The dependence of the chevron pattern on the modulation strength aligns well with simulations, confirming the accurate implementation of the CMCCD scheme.
We also evaluate the robustness of CMCCD, amplitude-modulated CCD (AMCCD) and phase-modulated CCD (PMCCD), by artificially adding detuning and Rabi static errors in a relatively low-noise environment, demonstrating increased robustness compared with the bare qubit.
The proposed CMCCD scheme offers a promising solution that combines precise qubit control with enhanced robustness in the presence of environmental noise and variability in qubits or driving fields across large-scale qubits.

\section{CCD by amplitude and phase modulation}
\label{sec:CCD}
\begin{figure}[hbt]
  \includegraphics[width=\columnwidth]{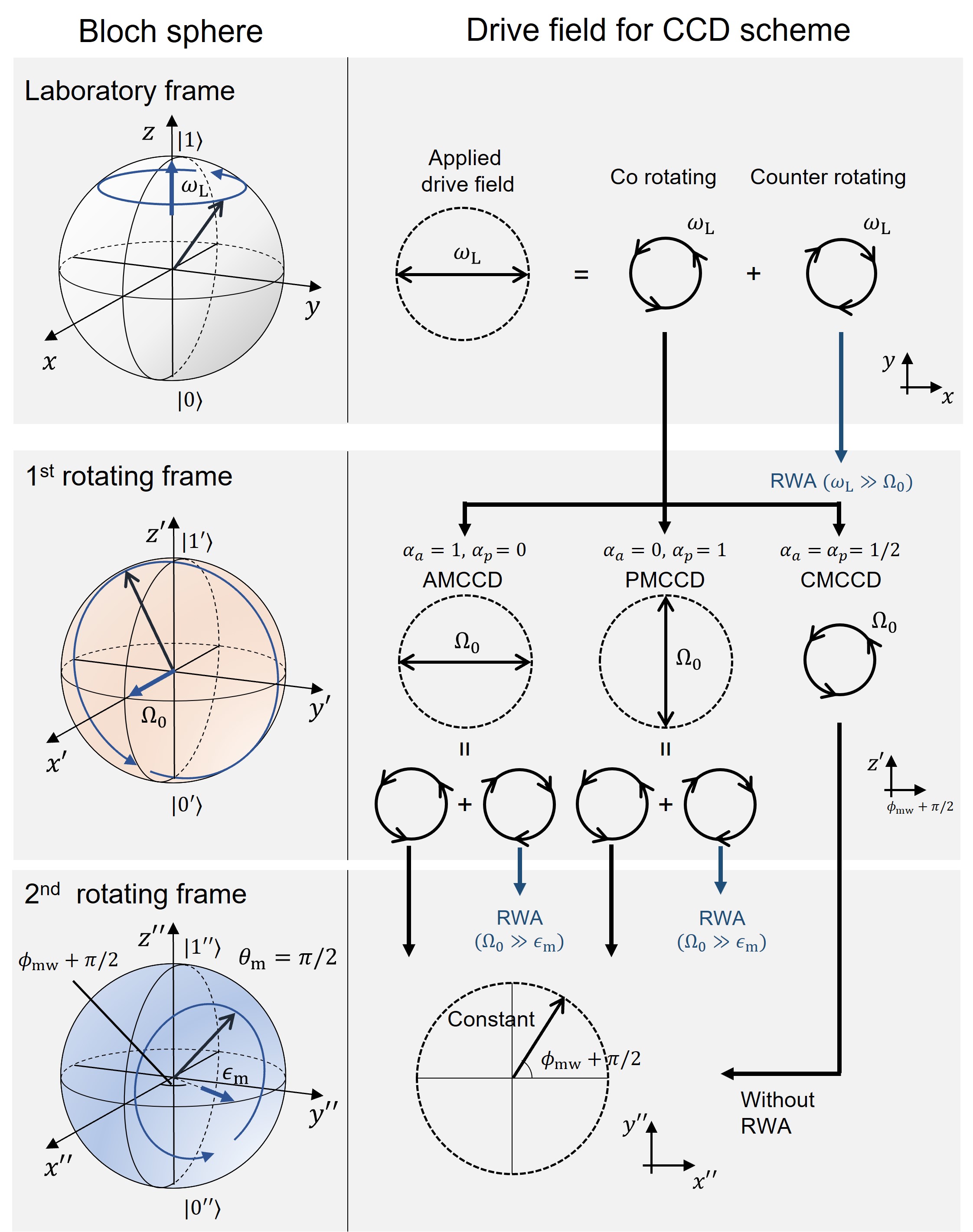}
  \caption{Polarization of drive field for different CCD schemes.
    To control the qubit, a drive field resonant with the Larmor frequency $\omega_{\rm{L}}$ is applied.
    In most cases, the applied drive field is linearly polarized, which is a superposition of corotating and counterrotating circularly polarized fields.
    In the first rotating frame, the counterrotating term results in a second harmonic field that can be neglected using RWA since $\Omega \ll \omega_L$.
    For a pure amplitude- or phase-modulated drive, in the first rotating frame the modulation results in a linearly polarized field along $x, y$, respectively. In the second rotating frame, the RWA is not so valid as typically  $\epsilon_{\rm{m}} < \Omega $, and the counterrotating term results in a systematic error in the gate. In this work, we propose CMCCD, where the amplitude and phase of the drive field are equally modulated,
    generating circular polarized field in the first rotating frame. Then in the second-rotating frame, there is no second harmonic term reducing the gate error.
  }
  \label{C0}
\end{figure}

Two primary methods of CCD protocols have been established: AMCCD and PMCCD.
Both approaches enable the realization of robust qubits; however, they suffer from a fundamental limitation—errors arising from the validity of the second RWA.
The electron spin-resonance mechanism is typically analyzed in the first rotating frame, where the corotating component of the linearly polarized MW drive $\omega_{\rm{mw}}$ and the Larmor precession $\omega_{\rm{L}}$ of the spin results in a dc drive field that drives a Rabi oscillation at frequency $\Omega_0$.
The counterrotating component gives rise to a second harmonic term at $2\omega_{\rm{mw}}$.
The effect of the second harmonic on the spin dynamics integrates to zero since $\Omega_0/2\pi\sim 1\text{-}10~\mathrm{MHz} \ll \omega_{\rm{mw}}/2\pi=10\text{-}20~\mathrm{GHz}$ in the typical spin-qubit situation, which is known as the RWA.

For CCD control, the MW drive is modulated, and in the first rotating frame, the modulation gives rise to an ac field of amplitude $\epsilon_{\rm{m}}$, and the Hamiltonian maps to the classic Rabi problem.
Hence, the analysis makes a second rotating frame transformation at the Rabi frequency $\Omega_0$.
However, the gate-speed is now determined by $\epsilon_{\rm{m}}$, and setting $\epsilon_{\rm{m}}\ll \Omega_0$ would result in very slow operations.
Hence, the non-negligible counterrotating term results in a systematic pulse-area error on resonance.
Since the RWA error stems from the use of linearly polarized driving fields, generating the circular polarization in the first rotating frame eliminates the counterrotating term.

Building on this insight, we introduce a generalized formulation of CCD drive as ($\hbar=1$)

\begin{widetext}
  \begin{align}
    H_{\rm{lab}}(t) & =\frac{\omega_{\rm{L}}}{2} \sigma_z +
    (\Omega_0+\Delta_\Omega)\left[\cos\left(\omega_{\rm{mw}}t+\phi_{\rm{mw}}-\frac{2\alpha_P\epsilon_{\rm{m}}}{\Omega_0}\sin(\Omega_0t-\theta_{\rm{m}})\right)\right.\notag \\
                    & \left.+ \frac{2\alpha_A\epsilon_{\rm{m}}}{\Omega_0}\sin(\Omega_0t
      -\theta_{\rm{m}})\sin\left(\omega_{\rm{mw}}t+\phi_{\rm{mw}}-\frac{2\alpha_P\epsilon_{\rm{m}}}{\Omega_0}\sin(\Omega_0
      t-\theta_{\rm{m}})\right)\right]\sigma_x,
  \end{align}
\end{widetext}
where $\sigma_i (i\in{\left\{x,y,z\right\}})$ are Pauli matrices, $\omega_{\rm{L}}$ is the qubit Larmor frequency, $\Omega_0$ is
the Rabi frequency, and $\Delta_\Omega$ is the Rabi frequency error (Rabi error), caused by MW fluctuations or
inhomogeneity.
The MW drive is characterized by its frequency $\omega_{\rm{mw}}/2\pi$ and phase $\phi_{\rm{mw}}$.
Modulation parameters of the CCD scheme are defined by modulation strength $\epsilon_{\rm{m}}$ and modulation phase
$\theta_{\rm{m}}$.
Here, we introduce two dimensionless parameters: amplitude-modulation ratio $\alpha_A$ and phase-modulation ratio
$\alpha_P$ ($\alpha_A,\alpha_P \geq  0$, $\alpha_A +\alpha_P = 1$).
For AMCCD ($\alpha_A=1,\alpha_P=0$) and PMCCD ($\alpha_A=0,\alpha_P=1$), and the bare qubit corresponds to $\alpha_A=\alpha_P=0$.
As discussed below, if amplitude and phase are equally modulated $\alpha_A=\alpha_P=1/2$, a circularly polarized drive field arises in the first rotating frame.
We name this scheme CMCCD.

To demonstrate this, we move to the first rotating frame defined by the time-dependent
Hamiltonian~\cite{wang2020coherence}
\begin{gather}
  H_0^{(1)}(t)=\left[\frac{\omega_{\rm{mw}}}{2}-\alpha_P\epsilon_{\rm{m}}\cos(\Omega_0t-\theta_{\rm{m}})\right]\sigma_z,
\end{gather}
the rotating frame Hamiltonian becomes

\begin{widetext}
  \begin{align}
    H_{\rm{rot}}^{(1)}(t) & =-H_0^{(1)}(t)+e^{i \int_{0}^{t} H_0^{(1)}(t') dt'}H_{\rm{lab}}(t)e^{-i \int_{0}^{t} H_0^{(1)}(t') dt'}\notag \\
                          & \approx\frac{\delta}{2}\sigma_z\notag + \frac{\Omega_0+\Delta_\Omega}{2}\sigma_{\phi_{\rm{mw}}} -
    \left(1+\frac{\Delta_\Omega}{\Omega_0}\right)\alpha_A\epsilon_{\rm{m}}\sin(\Omega_0t-\theta_{\rm{m}})\sigma_{\phi_{\rm{mw+\pi/2}}}
    +\alpha_P\epsilon_{\rm{m}}\cos(\Omega_0t-\theta_{\rm{m}})\sigma_z,
  \end{align}
\end{widetext}
where $\delta/2\pi = (\omega_{\rm{L}}-\omega_{\rm{mw}})/2\pi$ is detuning and we define
$\sigma_{\phi}\equiv\cos(\phi)\sigma_x+\sin(\phi)\sigma_y$.
Here, we use the first RWA since $\omega_{\rm{L}}/2\pi= 10\text{-}20~\mathrm{GHz} \gg\Omega_0/2\pi= 1\text{-}10~\mathrm{MHz}$.
The second term represents the standard Rabi drive field.
Importantly, the detuning term is orthogonal to the Rabi drive field, which suppresses dephasing and contributes to
the longer Rabi decay time $T_2^{\rm{Rabi}}$ than the dephasing time $T_2^*$~\cite{stano2022review}.
The third term corresponds to the ac drive field generated by the CCD modulation.
Notably, when $\alpha_A=\alpha_P=1/2$ (with zero Rabi error $\Delta_\Omega = 0$), the drive field becomes circularly
polarized (Fig.~\ref{C0}).
Then we move to the second rotating frame defined by $H_0^{(2)}=\frac{\Omega_0}{2}\sigma_{\phi_{\rm{mw}}}$ as
\begin{widetext}\label{Eq5}
  \begin{align}
    H_{\rm{rot}}^{(2)}(t)
                                                                                             & =\frac{\Delta_\Omega}{2}\sigma_{\phi_{\rm{mw}}} + \frac{\delta}{2} [\cos(\Omega_0 t)\sigma_z+\sin(\Omega_0
    t)\sigma_{\phi_{\rm{mw}}+\pi/2}]                                                         & \text{(Error terms)}\notag                                                                                      \\
                                                                                             & +\left[\alpha_P+\left(1+\frac{\Delta_\Omega}{\Omega_0}\right)\alpha_A\right]\frac{\epsilon_{\rm{m}}}{2}  \left[
    \cos(\theta_{\rm{m}})\sigma_z+\sin(\theta_{\rm{m}})\sigma_{\phi_{\rm{mw}}+\pi/2} \right] & \text{(Co-rotating term)} \notag                                                                                \\
                                                                                             & +\left[\alpha_P-\left(1+\frac{\Delta_\Omega}{\Omega_0}\right)\alpha_A\right]\frac{\epsilon_{\rm{m}}}{2}  \left[
      \cos(2\Omega_0 t-\theta_{\rm{m}})\sigma_z + \sin(2\Omega_0 t-\theta_{\rm{m}})\sigma_{\phi_{\rm{mw}}+\pi/2}
    \right]                                                                                  & \text{(Counter-rotating term)}
  \end{align}
\end{widetext}
The first line represents Rabi and detuning errors.
The corotating term yields a dc drive field for the qubit control at a frequency $\epsilon_{\rm{m}}/2\pi$, which is orthogonal to the Rabi error term and parallel to the oscillating detuning error.
If $\epsilon_{\rm{m}} \gg \Delta_{\Omega}$, a Rabi error tilts the control field, resulting in a rotation error that is quadratic, rather than linear in the Rabi error.
The oscillating detuning error modulates the precession frequency of the spin in the second rotating frame, leading to a phase shift that averages out to zero over time.
The counterrotating term is suppressed by RWA, which requires $\epsilon_{\rm{m}}\ll\Omega_0$, and slow operations.
For fast gates, the counterrotating term results in a residual pulse-area error on resonance.
However, by setting $\alpha_P=\alpha_A$, the counterrotating term is canceled and should lead to higher fidelity operations.

\begin{figure*}[htb]
  \includegraphics[width=\columnwidth*2]{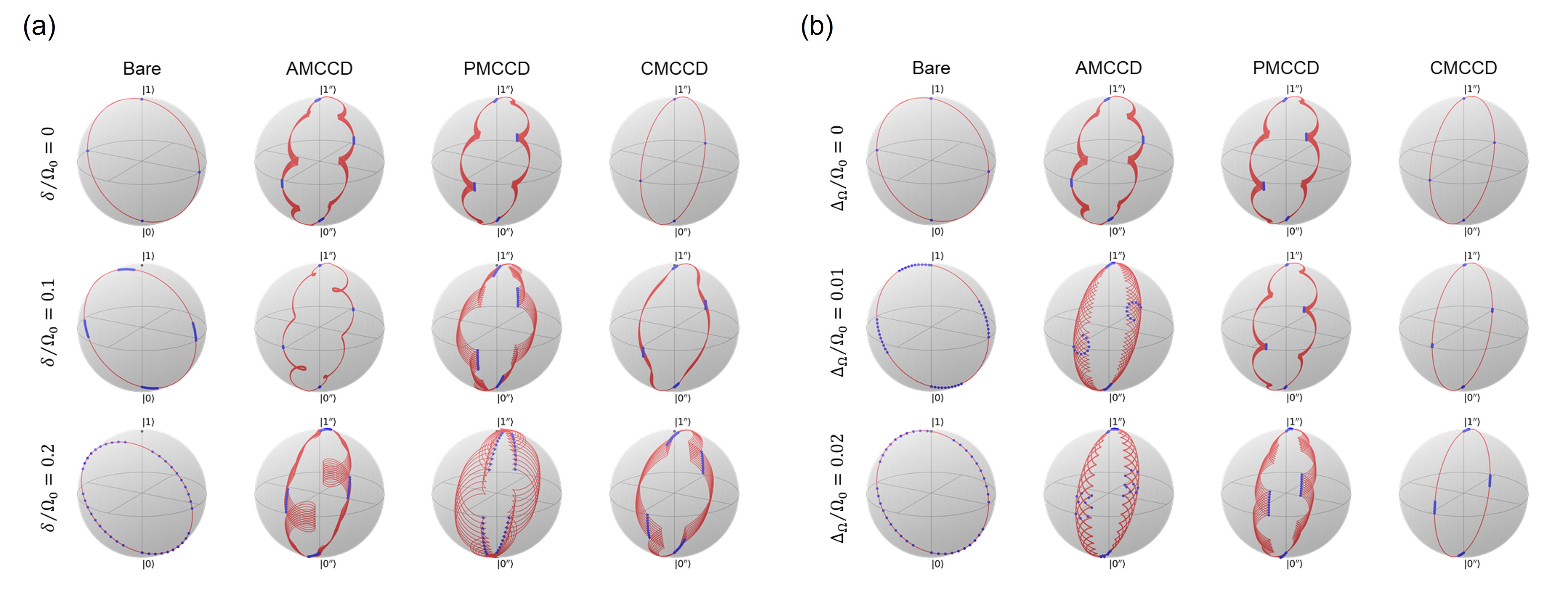}
  \caption{Quantum state trajectories of the bare qubit and CCD-protected qubits under (a) detuning error and (b)
    Rabi error.
    Modulation parameters are set to $\epsilon_{\rm{m}}=\Omega_0/4$, $\theta_{\rm{m}} = \pi/2$, and $\phi_{\rm{mw}} = 0$.
    Red lines trace the evolution of the quantum states over a total rotation angle of 20$\pi$, while blue markers
    indicate the positions corresponding to each $\pi/2$ rotation interval.
    Note that Bloch spheres for CCD-protected qubits are shown in the second rotating frame, while that for the bare
    qubit is shown in the first rotating frame. For the case of zero detuning or Rabi error, the AMCCD and PMCCD give a spread in the rotations resulting from integer $\pi/2$-gates, whereas both the bare qubit and CMCCD give perfect integer $\pi/2$-gates. When a static error is introduced, all CCD schemes have a low spread in the state-vector at integer $\pi/2$-gates, whereas the bare qubit suffers a large spread in outcomes.
  }
  \label{SIM1}
\end{figure*}

Figure~\ref{SIM1} illustrates the time evolution of the qubit under various CCD schemes and the bare qubit.
Here, we use modulation parameters as $\epsilon_{\rm{m}}=\Omega_0/4$, $\theta_{\rm{m}} = \pi/2$, and $\phi_{\rm{mw}} = 0$.
The red lines on the Bloch spheres trace the trajectories of quantum states with blue dots exhibiting the points of each $\pi/2$ pulse interval.
First, we compare the trajectories between CCD schemes at zero detuning [Fig.~\ref{SIM1}(a)].
AMCCD and PMCCD schemes exhibit distorted trajectories and the plotted blue dots are spread even in the case of zero detuning.
This distortion arises from the residual counterrotating term.
By contrast, the CMCCD scheme, which eliminates counterrotating terms, yields circular trajectories with accurately spaced $\pi/2$ rotations, indicating precise qubit control.
For the bare qubit, the trajectory remains circular. However, increasing detuning tilts the rotation axis,
resulting in a spread of the blue dots and increased gate error.
In CCD schemes, this spread is suppressed, thereby reducing error accumulation.
Figure~\ref{SIM1}(b) illustrates the robustness of each scheme against Rabi errors.
The bare qubit is highly sensitive to Rabi error, as indicated by widely spread blue dots.
This sensitivity is substantially reduced in all CCD schemes, highlighting their effectiveness in suppressing amplitude-induced errors.

\begin{figure}[htb]
  \includegraphics[width=\columnwidth]{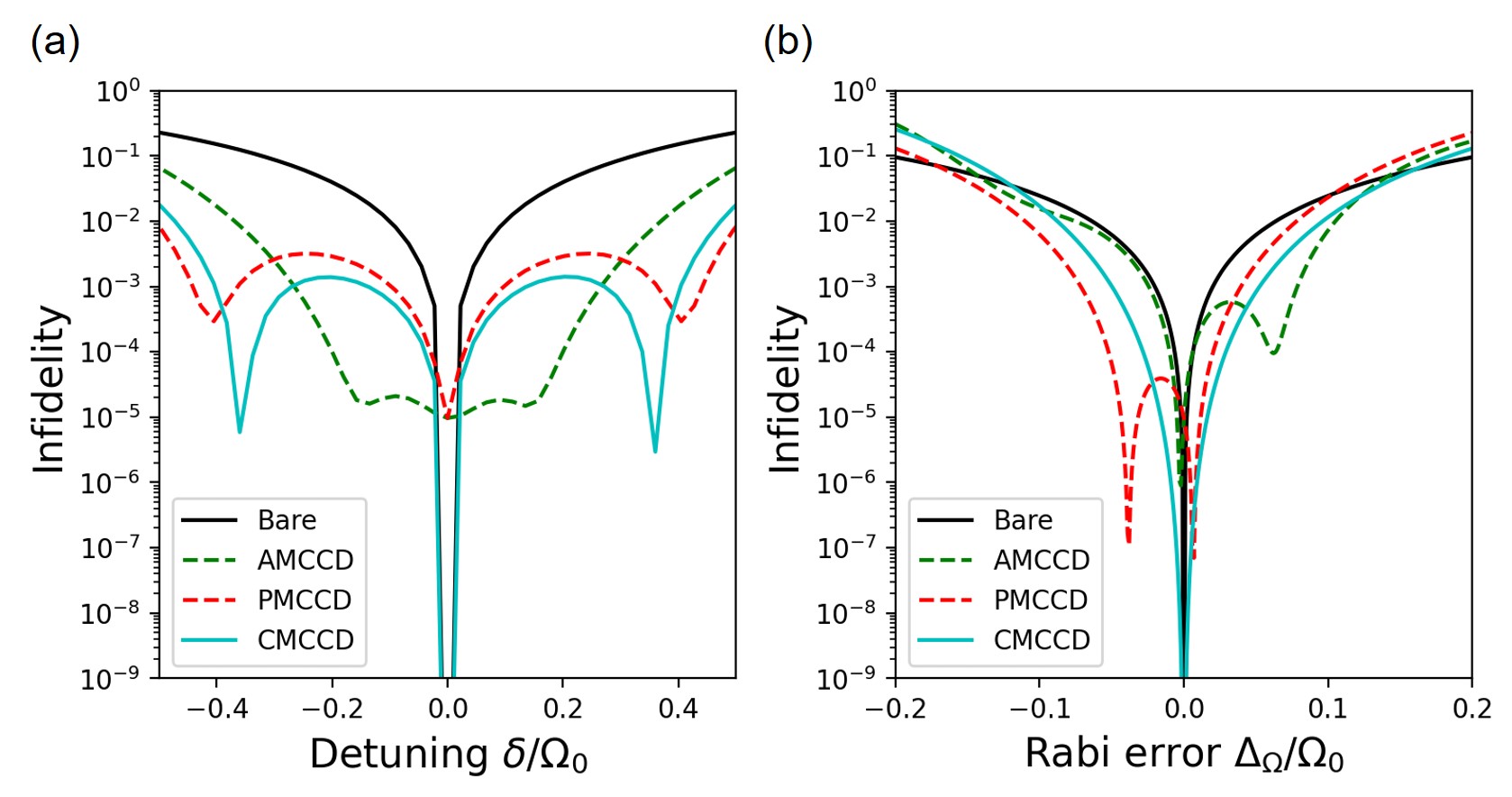}
  \caption{Comparison of state infidelity for the $Y_\pi$ gate as a function of (a) detuning error and (b) Rabi error. Detuning and Rabi errors are assumed to be quasistatic.
    Modulation parameters are set to $\epsilon_{\rm{m}}=\Omega_0/4$, $\theta_{\rm{m}} = \pi/2$, and $\phi_{\rm{mw}} = 0$.
    The $Y_\pi$ gate performs $Y$ axis rotation from $|0''\rangle$ to $|1''\rangle$ in the CCD schemes with a duration of $\pi/\epsilon_{\rm{m}}$ and from $|0'\rangle$ to $|1'\rangle$ in the bare qubit with a duration of $\pi/\Omega_0$.
    The state infidelity is defined as $1-|\langle 1''|Y_\pi|0''\rangle|^2 $ for CCD schemes and $1-|\langle 1'|Y_\pi|0'\rangle|^2$ for the bare qubit.
    The CMCCD combines the accuracy of a bare Rabi gate at low error with an improved tolerance to larger errors.
  }

  \label{SIM2}
\end{figure}

We further calculate the state infidelity for the $Y_\pi$ gate as a function of detuning and Rabi error (Fig.~\ref{SIM2}), using the same modulation parameters.
For bare qubits, the infidelity increases significantly with detuning or Rabi error, whereas CCD-protected qubits demonstrate strong resilience to both error sources.
While the degree of resilience varies among CCD implementations, CMCCD consistently achieves the highest gate fidelity in the low-error regime.
This variation in error sensitivity across schemes can be attributed to differences in how effectively counterrotating terms are suppressed.
The asymmetric dependence on Rabi error observed in AMCCD and PMCCD arises from residual counterrotating terms that are not entirely eliminated.
More detailed calculations are presented in Appendix~C.

\section{Device and experimental setup}
\label{sec:Device and experimental setup}
\begin{figure}[htb]
  \includegraphics[width=\columnwidth]{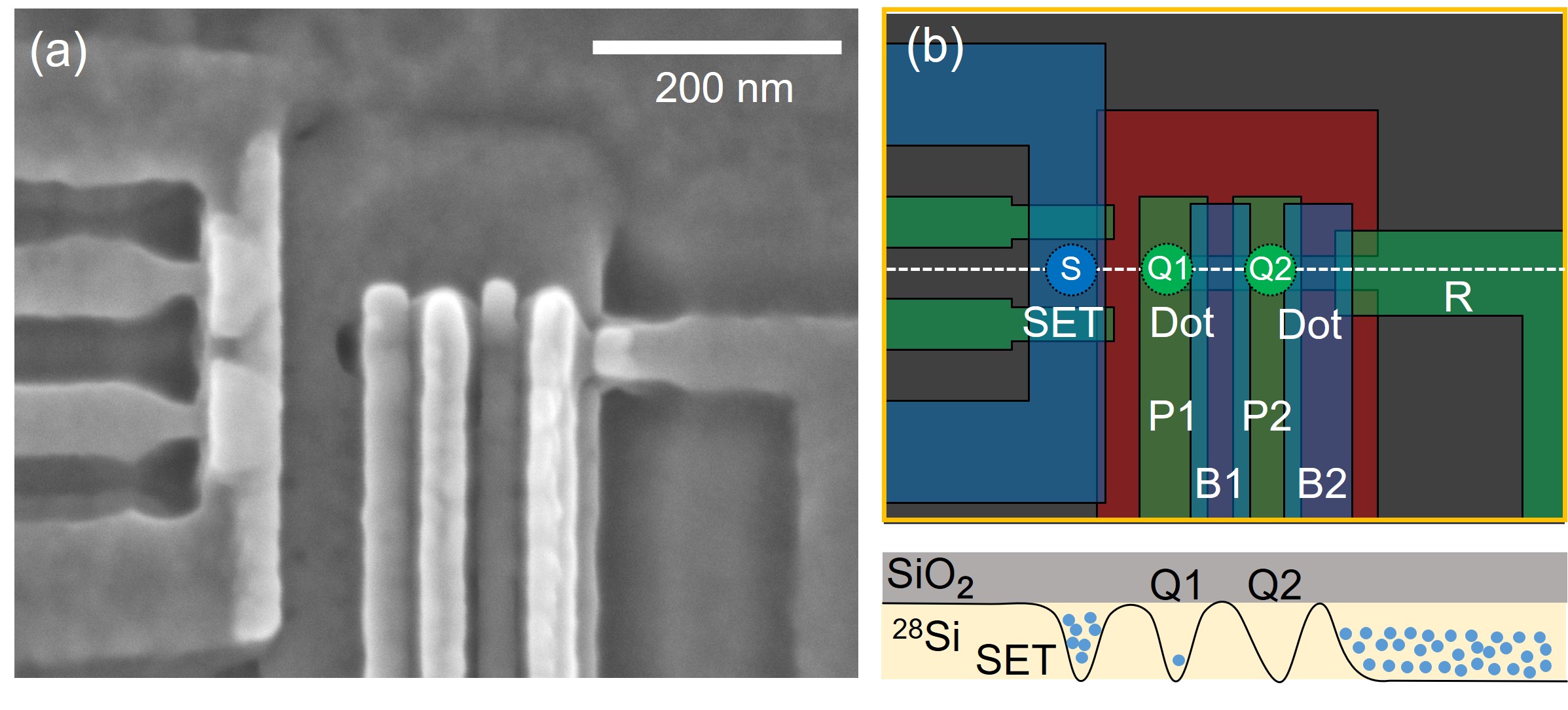}
  \caption{(a) The top-view scanning electron microscope (SEM) image and (b) its schematic illustration, including a cross-section view along the white dashed line.
    We operate the electron spin qubit confined in the left quantum dot (Q1).
  }
  \label{E0}
\end{figure}

To experimentally demonstrate the CCD scheme, we conduct our measurement on an isotopically purified SiMOS quantum dot~\cite{elsayed2024low}.
The device consists of two quantum dots where single electron spin qubits are accumulated and one single-electron transistor (SET) for charge sensing (Fig~\ref{E0}).
We readout the spin state via energy-selective tunneling~\cite{elzerman2004single}, detected through the SET sensor.
We apply static and pulsed voltages to the gates by voltage sources (QBLOX SPI D5a and Zurich Instruments HDAWG, respectively), while the SET current is measured using a current-voltage amplifier (Basel SP983c) and monitored in real time so that a single-electron tunneling event can be observed.
The spin-qubit manipulation is performed via the electron spin resonance (ESR), induced by a MW through an aluminum antenna.
The MW is generated by a MW source (Keysight PSG E8257D) and I/Q modulated by the signals from an arbitrary wave generator (Zurich Instruments HDAWG) to implement the CCD drive field.
The device is cooled down to a base temperature of about 10 mK using a dilution refrigerator.
Note that in this experiment, we use the electron spin qubit confined in the left quantum dot (Q1).
At a magnetic field of 0.7 T, the qubit has $T_2^*$=5.5 $\mathrm{\upmu s}$, $T_2^{\rm{Hahn}}$=69.1 $\mathrm{\upmu s}$, $T_2^{\rm{Rabi}}$ ($\Omega_0/2\pi$=1.2 MHz) =15.3 $\mathrm{\upmu s}$, see Fig.~\ref{S1} of Appendix~A. Compared with other reports in Si-28 devices, $T_2^{\rm{Hahn}}$ is relatively short~\cite{veldhorst2014addressable,zwerver2022qubits,steinacker2025industry}, indicating the presence of significant high-frequency noise.

\begin{figure*}[htb]
  \includegraphics[width=150mm]{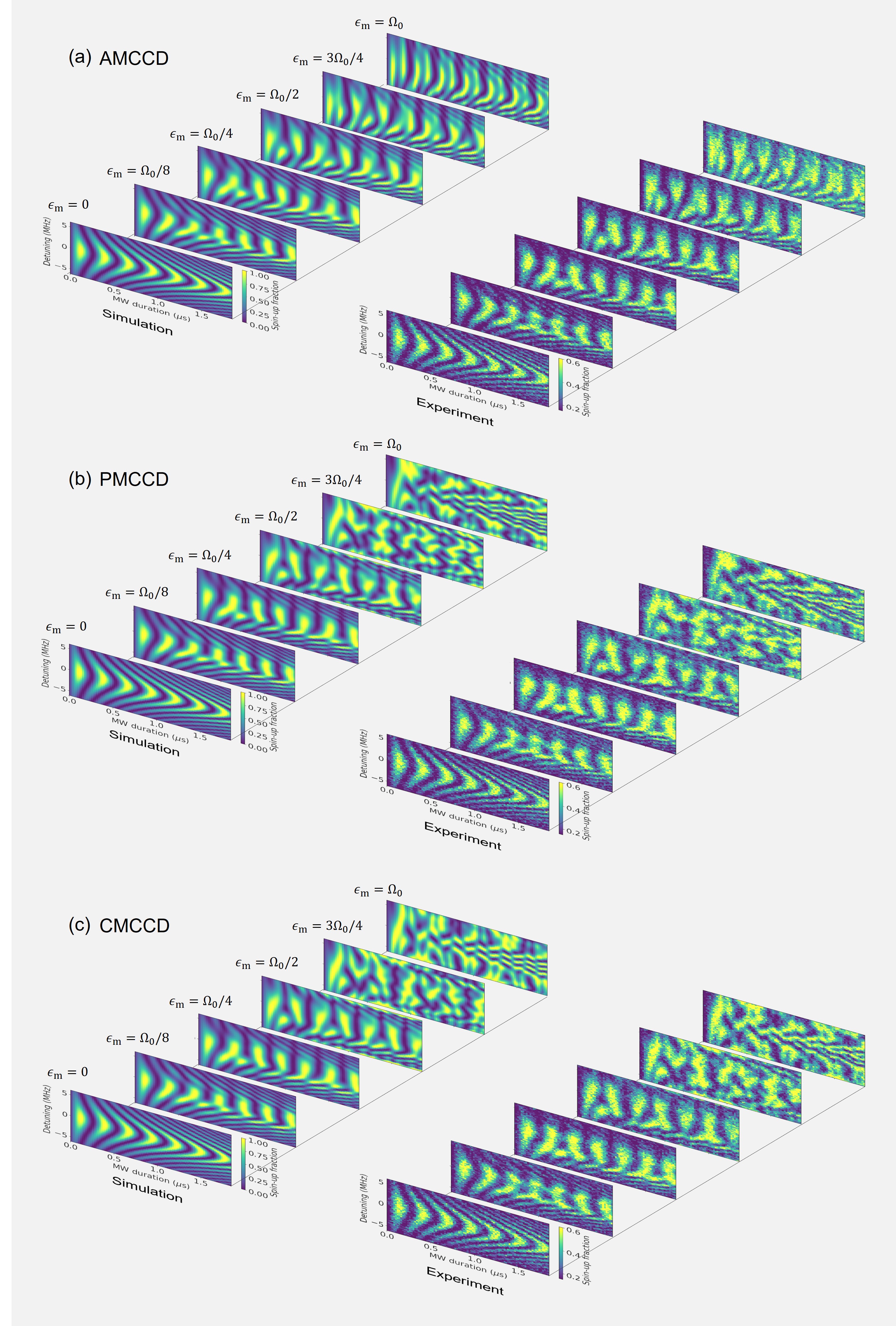}
  \caption{Chevron pattern of different CCD schemes, (a) AMCCD, (b) PMCCD, and (c) CMCCD measured at $\Omega_0 = 2\pi\times$ 3.6 MHz.
    The left panels show the simulation, and the right panels show the corresponding experimental results.
    We measure the chevron pattern by varying $\epsilon_{\rm{m}}$ from 0 to $\Omega_0$.
    When  $\epsilon_{\rm{m}} = 0$, the condition is the same as the bare qubit.
    As $\epsilon_{\rm{m}}$ increases, the chevron pattern gradually transforms into a ladderlike structure, indicating enhanced robustness against detuning.
    With further increase toward $\Omega_0$, the ladderlike shape begins to distort.
  }
  \label{E1}
\end{figure*}

\clearpage
\newpage
\section{Properties of stabilized Rabi oscillation}
\label{sec:experimental results1}

In this section, we investigate the properties of stabilized Rabi oscillation by different CCD schemes.
As explained in the previous section, the CCD scheme can improve the robustness of both detuning error and Rabi error.
First, to demonstrate the amplitude-phase-modulated CCD, we show simulation and experimental results of Rabi oscillations over the detuning in Fig.~\ref{E1}.
We measure the Rabi oscillations by three types of CCD schemes at the bare qubit Rabi frequency $\Omega_0/2\pi$ = 3.6 MHz.
From the usual Rabi chevron pattern ($\epsilon_{\rm{m}} = 0$), with increasing $\epsilon_{\rm{m}}$, the chevron pattern is modified and becomes a more complex pattern.
When $\epsilon_{\rm{m}}$ is small, there is almost no difference in the maps between the three CCD schemes, but as $\epsilon_{\rm{m}}$ increases, the difference becomes more pronounced.
Simulation and experimental results are in good agreement, demonstrating the proper implementation of the amplitude-phase modulated CCD schemes.
Note that the slight difference between the simulation and experimental results of AMCCD [Fig.~\ref{E1}(a)] for large $\epsilon_{\rm{m}}$ is due to the nonlinearity of the MW amplifier.
With a view to gate operations, we use a modulation strength of $\epsilon_{\rm{m}}=\Omega_0/4$. For the MW power used, the overall power remains within the linear response regime of the amplifier.

\begin{figure}[htb]
  \includegraphics[width=\columnwidth]{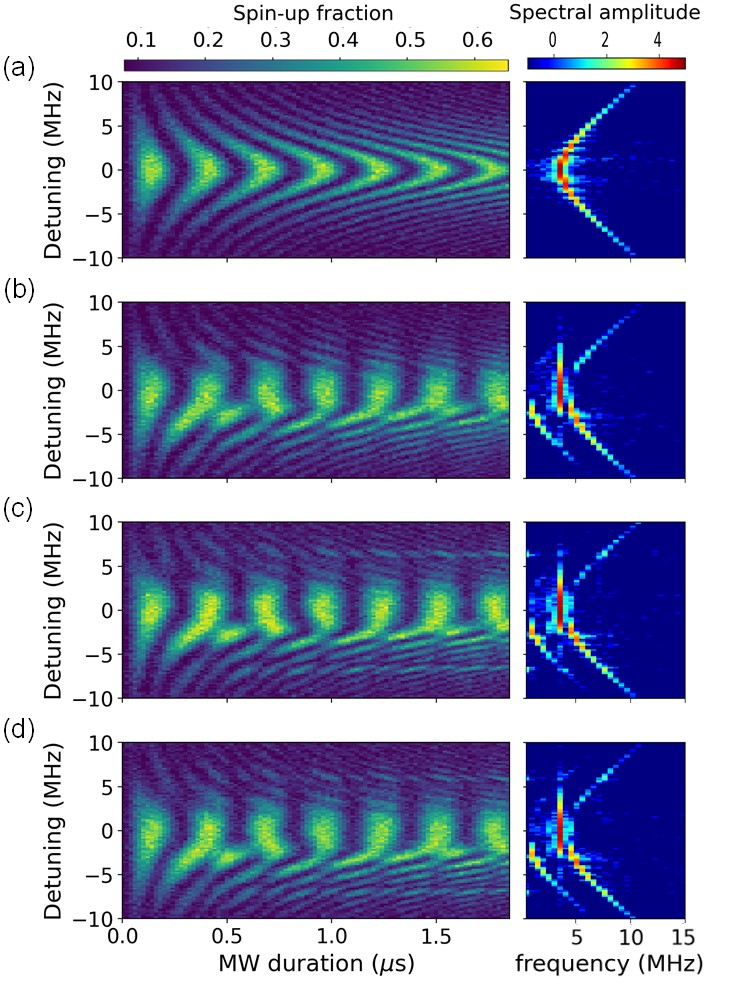}
  \caption{Comparison of Rabi chevrons [spin up fraction following drive vs duration and detuning of the drive (left panels), and their Fourier transform (right panels)].
    (a) Bare qubit, (b) AMCCD, (c) PMCCD and (d) CMCCD.
    We measure the chevron patterns under the same conditions, where $\Omega_0 = 2\pi\times3.6$ MHz for the bare
    qubit, and $\epsilon_{\rm{m}} = \Omega_0/4 = 2\pi\times0.9$ MHz for the CCD schemes.
    The bare qubit exhibits a classic chevron pattern with an effective Rabi frequency that has hyperbolic dependence on detuning. The CCD patterns have a ladder-like structure. In the Fourier domain, near zero detuning there is a single frequency component at the modulation frequency, which is robust against detuning errors.
  }
  \label{E2}
\end{figure}

For the bare qubit with a detuning $\delta$, the spin rotates about a tilted axis at the oscillation frequency as $\sqrt{\Omega_0^2+\delta^2}$, revealing the usual chevron pattern [Fig.~\ref{E2}(a)].
In the CCD schemes, the qubit is rotating about two axes for executing the dynamic decoupling, decreasing the sensitivity to the detuning error.
The Rabi oscillations over the detuning in CCD schemes exhibit an improvement in robustness against the detuning error.
The chevron patterns of the bare qubit are changed to ladderlike structures as shown in Figs.~\ref{E2}(b)-(d).
The robustness of the CCD against the detuning is also clearly evident in the Fourier spectra.
In contrast with the hyperbolic dependence of the detuning for the bare qubit, the spectra in CCD schemes are flat in a certain range of the detuning.
For silicon spin qubits, the detuning error varies in time due to fluctuations of nuclear spins from $^{29}$Si isotopes~\cite{khaetskii2002electron,hanson2007spins,assali2011hyperfine}.
In addition, when manipulating multi qubits, variations in the $g$ factor cause differences in the resonance frequency between the qubits~\cite{ferdous2018interface,cifuentes2024bounds}, resulting in detuning errors.
These results show that CCD schemes can perform qubit operations that are robust against variations in the detuning.

\begin{figure}[htb]
  \includegraphics[width=\columnwidth]{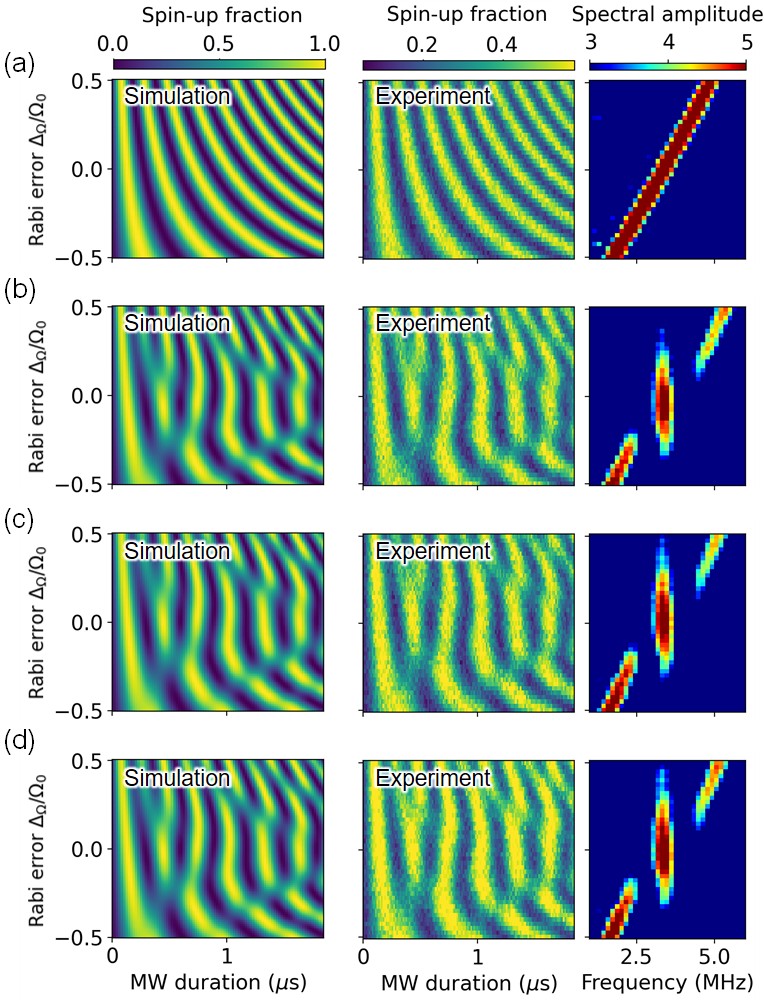}
  \caption{Sensitivity of CCD to static Rabi errors. Dependence of Rabi oscillations on static Rabi errors measured at $\Omega_0 = 2\pi\times3.3$ MHz for the bare
    qubit and $\epsilon_{\rm{m}} = \Omega_0/4 = 2\pi\times0.825$ MHz for the CCD schemes.
    (a) Bare qubit, (b) AMCCD, (c) PMCCD and (d) CMCCD.
    The left panels show simulation results.
    The middle panels are experimental results, showing excellent agreement with simulations.
    The right panels show the Fourier spectrums over the MW duration time.
    The bare qubit has a linear dependence of the Rabi frequency on the Rabi error.
    By applying the CCD scheme, the Rabi frequency becomes constant within a certain range of Rabi errors, as can
    be seen from the Fourier spectra, showing the reduction in the sensitivity to Rabi error.
  }
  \label{E3}
\end{figure}

Next, we investigate robustness against the Rabi error, which is caused by the fluctuations in the qubit drive field.
When considering the multiple silicon spin qubits, the Rabi error can also arise from variations in the $g$ factor and the spatial inhomogeneity of the MW.
In this experiment, we artificially increase or decrease the MW amplitude and evaluate the robustness against a static Rabi error.
Figures~\ref{E3}(a)-(d) show the simulation and experimental results of Rabi oscillations over the Rabi error.
For the bare qubit, as the Rabi error increases, the Rabi frequency becomes higher, and this is expressed by the linear dependence of the Fourier spectrum [Fig.~\ref{E3}(a)].
This linear dependence shows that the bare qubit is vulnerable to the Rabi error.
By applying CCD schemes, within a range of around $\pm20\%$ of the Rabi error, the obtained Rabi frequency is
insensitive to the Rabi error as shown in Figs.~\ref{E3}(b)-(d).
These maps of Rabi error dependence of Rabi oscillations are in excellent agreement with those predicted from
simulations.

From the above examinations, we demonstrate that CCD schemes are resistant to detuning and Rabi errors.
Furthermore, we confirm that the proposed CMCCD achieves the same effect as the conventional single-amplitude/phase-
modulation CCD and is consistent with the simulation.
In the next section, we will evaluate the performance of the gate fidelity for CCD qubits via a randomized
benchmarking method.

\begin{figure}[htb]
  \includegraphics[width=\columnwidth]{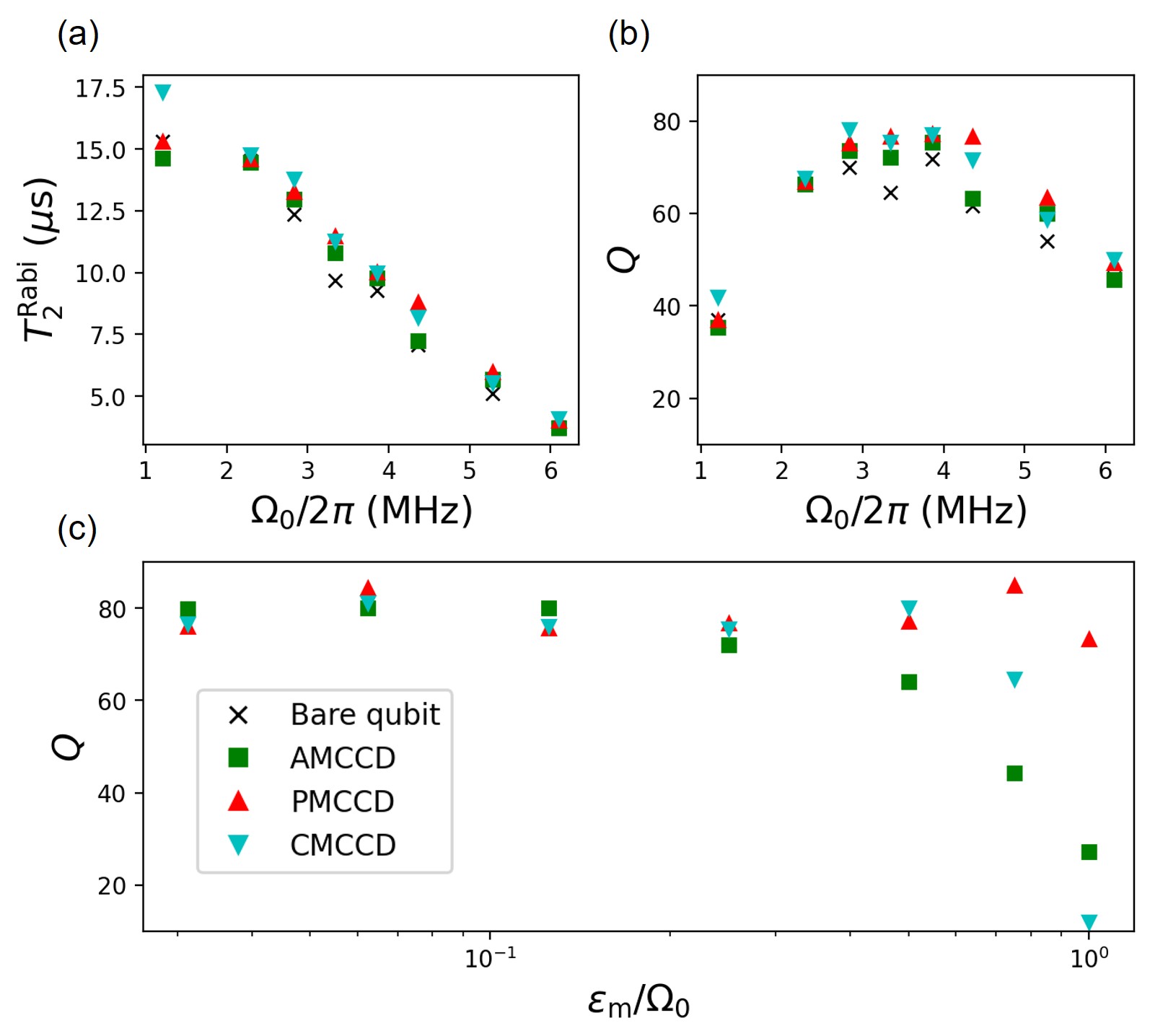}
  \caption{ Properties of Rabi oscillations in the laboratory frame without the artificial error.
  (a) $\Omega_0/2\pi$ dependence of $T_2^{\rm{Rabi}}$. $T_2^{\rm{Rabi}}$ decreases as $\Omega_0/2\pi$
  increases due to the increase in the heating from the MW pulse.
  (b) $\Omega_0/2\pi$ dependence of quality factor $Q$. No significant improvement is found by using CCD
  schemes.
  (c) $\epsilon_{\rm{m}}$ dependence of quality factor $Q$ measured at $\Omega_0=2\pi\times3.6$ MHz.
  In PMCCD, there is no significant dependence on $\epsilon_{\rm{m}}$.
  In AMCCD and CMCCD, increasing $\epsilon_{\rm{m}}$ decreases $Q$ due to the MW induced noise caused by the amplitude modulation.
  }
  \label{E4}
\end{figure}

At the end of this section, we examine the characteristics of the CCD schemes for Rabi oscillations with no
artificial errors.
Figures~\ref{E4}(a) and \ref{E4}(b) show Rabi frequency $\Omega_0/2\pi$ dependence of Rabi oscillation decay time
$T_2^{\rm{Rabi}}$ and quality factor $Q$.
Here, we evaluate the coherence protection effect using the Rabi oscillations measured in the laboratory frame.
We define $Q$ as the decay time $T_2^{\rm{Rabi}}$ divided by the $\pi$ rotation time $T_{\pi}$, $Q =T_2^{\rm{Rabi}}/T_{\pi}$.
The $\pi$ rotation time is determined from the conventional Rabi frequency in the laboratory frame ($T_\pi = \pi / \Omega_0$).
We use $\epsilon_{\rm{m}} = \Omega_0/4$ for CCD schemes and extract the decay time $T_2^{\rm{Rabi}}$ by fitting to an exponentially decaying sinusoidal.
For comparison, we plot the bare qubit results.
As shown in Fig.~\ref{E4}(a), the decay time $T_2^{\rm{Rabi}}$ decreases as $\Omega_0/2\pi$ increases.
This decrease is probably due to the MW-induced noise observed in the ESR or EDSR experiments~\cite{undseth2023hotter}.
The maximum $Q$ of the bare qubit is about 70, but this value is similar to those of three CCD schemes as
shown in Fig.~\ref{E4}(b).
To further investigate the CCD schemes, we plot the quality factor $Q$ as a function of modulation strength
$\epsilon_{\rm{m}}$ [Fig.~\ref{E4}(c)].
When $\epsilon_{\rm{m}}$ is small, no significant differences are observed between the CCD schemes, but as $\epsilon_{\rm{m}}/\Omega_0$ approaches 1, $Q$ decreases in AMCCD and CMCCD.
A decrease in $Q$ factor with $\epsilon_{\rm{m}}$ is expected due to the increasing strength of the counterrotating term in the second rotating frame. However, a decrease in $Q$ is also observed for CMCCD where this error is suppressed, suggesting that the higher MW power, see Appendix~B, leads to additional dephasing

We note here that the CCD schemes cannot improve the quality $Q$, unlike the previous reports from using a natural
silicon quantum dot~\cite{kuno2026robust}.
This observation indicates that the improvement of $Q$ is limited by noise components that are not suppressed by the CCD schemes. Since CCD primarily suppresses low-frequency fluctuations, the remaining higher-frequency noise likely limits the achievable $Q$.

\clearpage
\newpage
\section{Characteristics of CCD-protected qubit}
\label{sec:experimental results2}

\begin{figure}[htb]
  \includegraphics[width=\columnwidth]{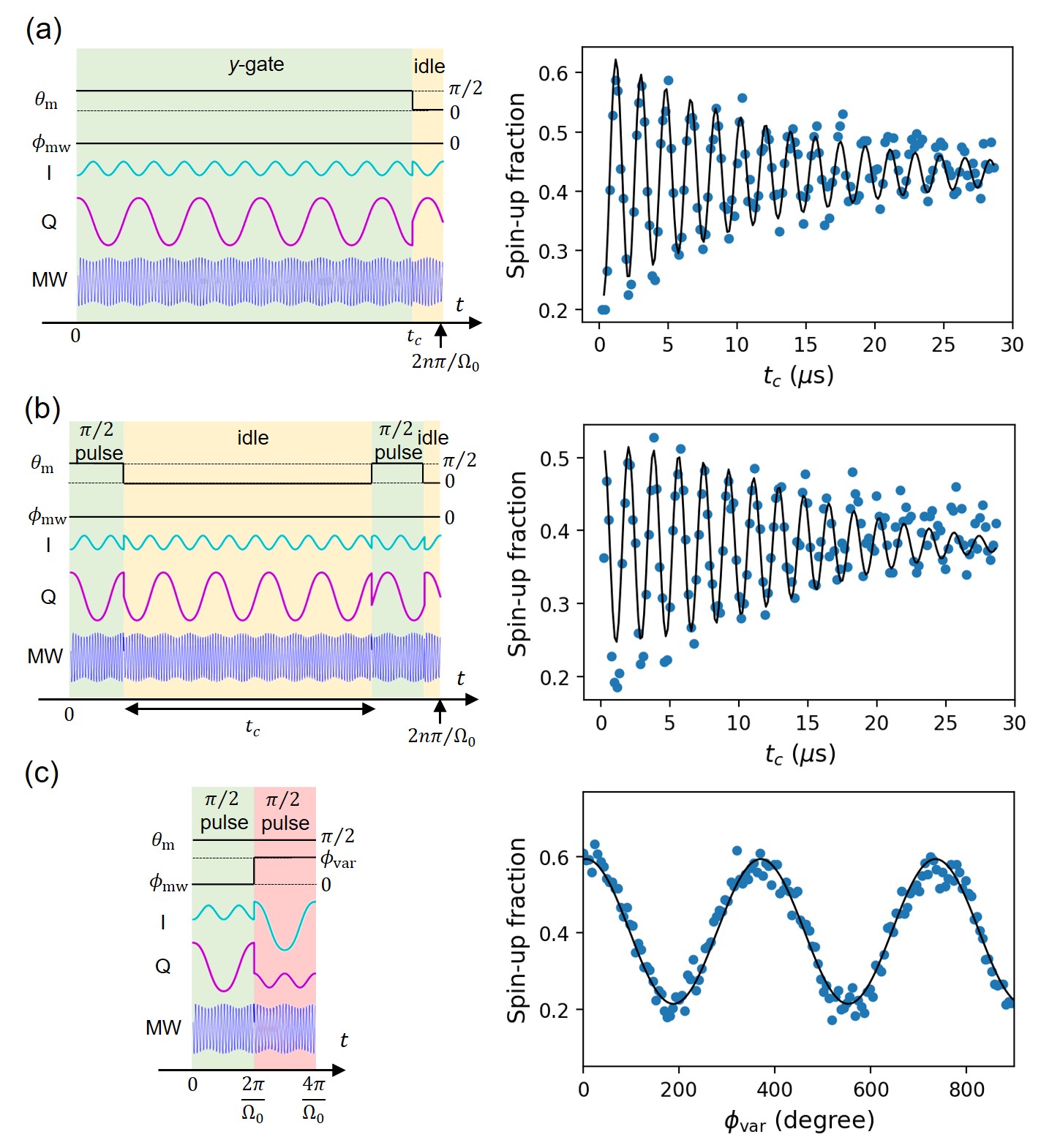}
  \caption{ Control of the double-dressed qubit by CMCCD scheme.
  The left schematics of each figure show the pulse sequences.
  All experiments are measured at $\Omega_0 =2\pi\times2.2$ MHz and $\epsilon_{\rm{m}} = \Omega_0/4$.
  (a) CCD-Rabi experiment. The first pulse executes $y$-axis rotation of $t_c$ duration.
  The second pulse is an idle pulse to adjust the readout time to $2n\pi/\Omega_0$.
  This matches the laboratory and the second rotating frame basis for readout~\cite{kuno2026robust}.
  The decay time $T_2^{\rm{CCD\text{-}Rabi}} = 12.8$ $\mathrm{\upmu s}$.
  (b) CCD-Ramsey experiment. The idle pulse of duration $t_c$ is applied between two $\pi/2$ pulses, followed
  by the idle pulse for the readout matching. The decay time $T_2^{\rm{CCD\text{-}Ramsey}} = 17.7$ $\mathrm{\upmu s}$.
  (c) Two-axis control experiment. We apply two consecutive $\pi/2$ pulses and vary the MW phase
  $\phi_{\rm{var}}$.
  The spin-up probability as a function of $\phi_{\rm{var}}$ oscillates with a period of 2$\pi$,
  demonstrating the control of the rotation axis in the double-dressed qubit by CMCCD.
  }
  \label{E5}
\end{figure}

In the previous section, we focused on the Rabi oscillations in the first rotating frame.
This section examines the CCD-protected qubit defined in the double-dressed state.
To take advantage of the robustness provided by CCD schemes, we need to control the CCD-protected qubit defined in the second rotating frame.
The readout and control method of the CCD-protected qubit was discussed in Ref~\cite{kuno2026robust}.
In this study, we first confirm the qubit control using the  CMCCD scheme.
As described in Eq.~(3), we achieve $z$ rotation by setting $\theta_{\rm{m}} = 0$ for idle pulse, and $x$-$y$ plane axis rotation for the gate operation by setting $\theta_{\rm{m}} = \pi/2$, varying the rotation axis with $\phi_{\rm{mw}}$.
Figures~\ref{E5}(a)-(c) show basic control pulse sequences and experimental results.
As shown in Fig.~\ref{E5}(a), to demonstrate the $y$-gate operation, we apply pulse duration of $t_c$, where
$\theta_{\rm{m}} = \pi/2$ and $\phi_{\rm{mw}} = 0$, followed by the idle pulse for readout matching.
We observe the Rabi oscillation in the second rotating frame at the frequency of $\epsilon_{\rm{m}}/2\pi$ as expected.
Next, we measure the coherence time of the double dressed qubit by CMCCD scheme as shown in Fig~\ref{E5}(b).
The observed oscillations correspond to the $z$ rotation of the idle pulse.
In the experiments, we evaluate the decay times by fitting to a decaying exponential and obtain
$T_2^{\rm{CCD\text{-}Rabi}} = 12.8$ $\mathrm{\upmu s}$ and $T_2^{\rm{CCD\text{-}Ramsey}} = 17.7$ $\mathrm{\upmu s}$.
In Fig.~\ref{E5}(c), we then demonstrate the two-axis control by the MW phase $\phi_{\rm{mw}}$.
After a $\pi/2$ pulse of the MW phase $\phi_{\rm{mw}}=0$, we apply the additional $\pi/2$ pulse, varying the MW
phase of $\phi_{\rm{mw}}=\phi_{\rm{var}}$.
The obtained spin-up fraction oscillates with a period of 2$\pi$, thus achieving the two-axis control in the CMCCD
scheme.

\begin{figure}[htb]
  \includegraphics[width=\columnwidth]{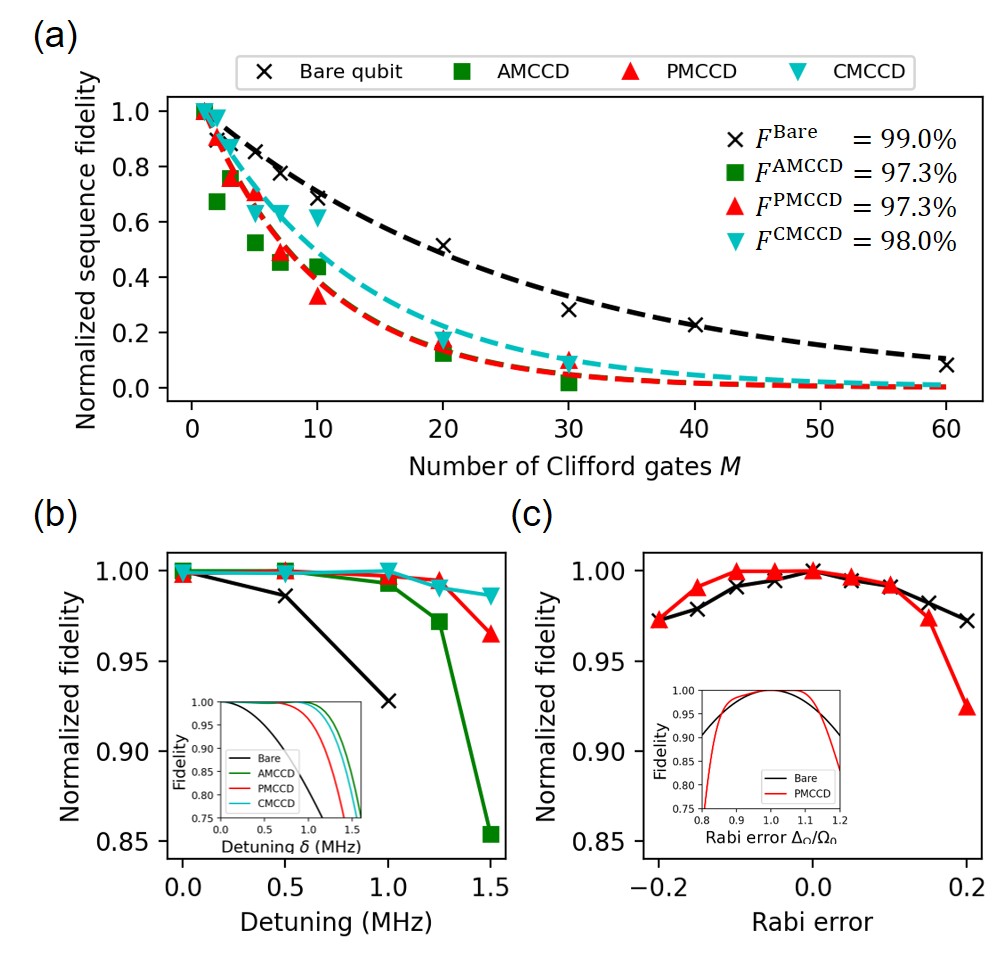}
  \caption{ Robustness of gate fidelity using randomized benchmarking.
    Experiments are conducted at $\Omega_0 =2\pi\times2.2$ MHz and $\epsilon_{\rm{m}} = \Omega_0/4$.
    (a) Randomized benchmarking of the bare qubit and CCD qubits. The sequence is repeated for $K$=15 random
    Clifford gate sets.
    The normalized gate fidelity as a function of (b) detuning and (c) Rabi error.
    For comparison, we put insets showing the simulation results of $Y_\pi$ gate-state fidelity, the same data
    as in Fig.3.
  }
  \label{E6}
\end{figure}

Finally, we investigate the gate fidelity and the robustness of three CCD schemes and compare them with the bare qubit.
In the following experiments, for CCD schemes, we use the modulation strength $\epsilon_{\rm{m}} = \Omega_0/4$ to minimize the counterrotating term appearing in the AMCCD and PMCCD schemes~\cite{kuno2026robust}.
This choice also ensures that Clifford gates can be implemented within an integer number of periods of the modulation frequency, which simplifies the experimental implementation and readout.
Figure~\ref{E6}(a) shows the reference randomized benchmarking measurement data to determine the average single gate fidelity~\cite{knill2008randomized,muhonen2015quantifying}.
In this sequence, $M$ random Clifford gates are applied, followed by a recovery Clifford gate that returns the final spin state to spin-up or spin-down.
We plot the differences in spin-up fraction between two cases and fit the data to an exponential curve of the form $(2F_{\rm{c}}-1)^M$.
From this fit, we extract the average Clifford gate fidelity $F_c$. The fidelity of an average single gate, denoted as $F$, is then calculated using the relation $F = 1-(1-F_{\rm{c}})/1.875$~\cite{muhonen2015quantifying}.
The fidelity is lower with CCD schemes than the bare qubit.
As can be inferred from the results in the previous section that the dominant noise in this experimental system cannot be removed by the CCD, so the fidelity is reduced simply by slowing the gate speed by a factor of 4 ($\epsilon_{\rm{m}} = \Omega_0/4$).
On the other hand, the CCD scheme can improve robustness against the detuning and the Rabi errors.
To characterize the robustness of CCD qubits, we artificially apply static detuning or Rabi error and evaluate the gate fidelity.
Figure~\ref{E6}(b) shows the measured gate fidelity as a function of detuning.
The fidelity of the bare qubit decreases as detuning increases, whereas CCD qubits show a decrease in sensitivity to detuning, enabling the fidelity to be maintained even under some detuning.
Figure~\ref{E6}(c) presents the Rabi error dependence of the fidelity.
The effect of PMCCD is particularly evident in the Rabi error $\Delta_\Omega/\Omega_0$ range of $-$0.15 to 0 [Fig.~\ref{E6}(c)].
However, the improvement in robustness against Rabi errors is not as significant as that against detuning errors.
This is because CCD qubits have a double dynamical decoupling effect against detuning errors, introduced by the two frame transforms.
Note that as shown in Fig.~{\ref{SIM2}}(b), simulations suggest that PMCCD, which operates without amplitude modulation, offers better robustness to Rabi errors and allows easier experimental comparison.
Based on this, we used PMCCD to study the effect of the Rabi error.

\section{Summary}
\label{sec:summary}
In this study, we explore the implementation of the CCD scheme for qubit applications.
A limitation of the conventional CCD scheme lies in the RWA in the second rotating frame, which can degrade the gate fidelity.
To address this issue, we propose the CMCCD scheme, which employs dual modulation of MW amplitude and phase for qubit driving.
This approach eliminates the residual counterrotating term and dynamically decouples qubits from noise, thereby obviating the need for RWA.

We experimentally implement the CMCCD scheme using an electron spin-qubit in an isotopically purified silicon quantum dot.
The dependence of the chevron pattern on the CCD modulation strength $\epsilon_{\rm{m}}$ aligns well with simulation results, confirming the accurate implementation of the proposed scheme.
We further assess the robustness against detuning and Rabi errors by measuring Rabi oscillations and their Fourier spectrum over MW duration.
Our findings indicate a significant improvement in robustness compared with bare qubits, consistent with simulation predictions.
Moreover, this enhanced robustness extends to CCD qubits defined in the double-dressed state.
We evaluate the effectiveness of the CCD schemes by measuring gate fidelity through randomized benchmarking under artificially induced errors.
The robustness is significantly improved with CCD schemes, demonstrating advantages for stable qubit control under imperfect and noisy conditions.
We note that while the robustness is improved, the base gate fidelity is not improved using the CCD schemes, unlike the previous study in a natural silicon device with higher low frequency noise~\cite{kuno2026robust}.
In our device, the measured $T_2^*$ = 5.5 $\mathrm{\upmu s}$ corresponds to a qubit frequency fluctuation of approximately $\sigma/2\pi \approx 40\,\mathrm{kHz}$, using the relation $T_2^* = \sqrt{2}/\sigma$.
In the experiments for evaluating the gate fidelity in Fig.~\ref{E6}, we use a Rabi frequency of 2.2 MHz.
This gives $\sigma/\Omega_0$ = 0.02, which by inspection of Fig.~\ref{SIM2}(a), indicates that this device is too quiet to benefit from CCD control.
However, in previous work on a natural silicon device, where $\sigma/\Omega_0$ = 1.7 ($T^*_2$=143 ns, $\Omega_0/2\pi$=940 kHz), the improvement can be dramatic~\cite{kuno2026robust}.

To place the present results in the broader context of error-mitigation strategies, we briefly compare the CCD approach with conventional pulse-shaping techniques~\cite{yang2019silicon,xue2022quantum,rimbach2023simple}. Pulse shaping mitigates control errors by tailoring the temporal envelope of individual pulses, for example, using smooth ramping functions, and more generally through waveform optimization, including both analytical designs and numerical optimal control. In contrast, CCD achieves robustness through Hamiltonian engineering in the dressed-state picture, thereby protecting the qubit during the operation. While pulse-shaping methods often rely on waveform optimization tailored to a specific system and, in some cases, to a noise model, CCD provides robustness through its intrinsic dynamical structure under continuous driving. This continuous protection may be advantageous in situations where noise characteristics are not fully known, and may also help maintain qubit coherence during idle periods.

The CCD scheme presents a valuable tool for controlling qubits in noisy environments and under variations in qubit characteristics or driving fields.
This method provides reliable control strategies indispensable for multi qubit systems in fault-tolerant quantum computers.
Furthermore, as the proposed CMCCD scheme does not require RWA, it is anticipated to enable higher fidelity gate operations applicable not only to silicon spin qubits but also other types of qubits, such as trapped atoms~\cite{nunnerich2025fast}, cold atoms~\cite{wang2025individual}, superconducting qubits, and NV centers~\cite{cai2012robust}.

\section*{Acknowledgements}
This work was supported by JST Moonshot R $\&$ D Grant No. JPMJMS2065.

\section*{APPENDIX A: Bare qubit properties}

\begin{figure}[htb]
  \includegraphics[width=\columnwidth]{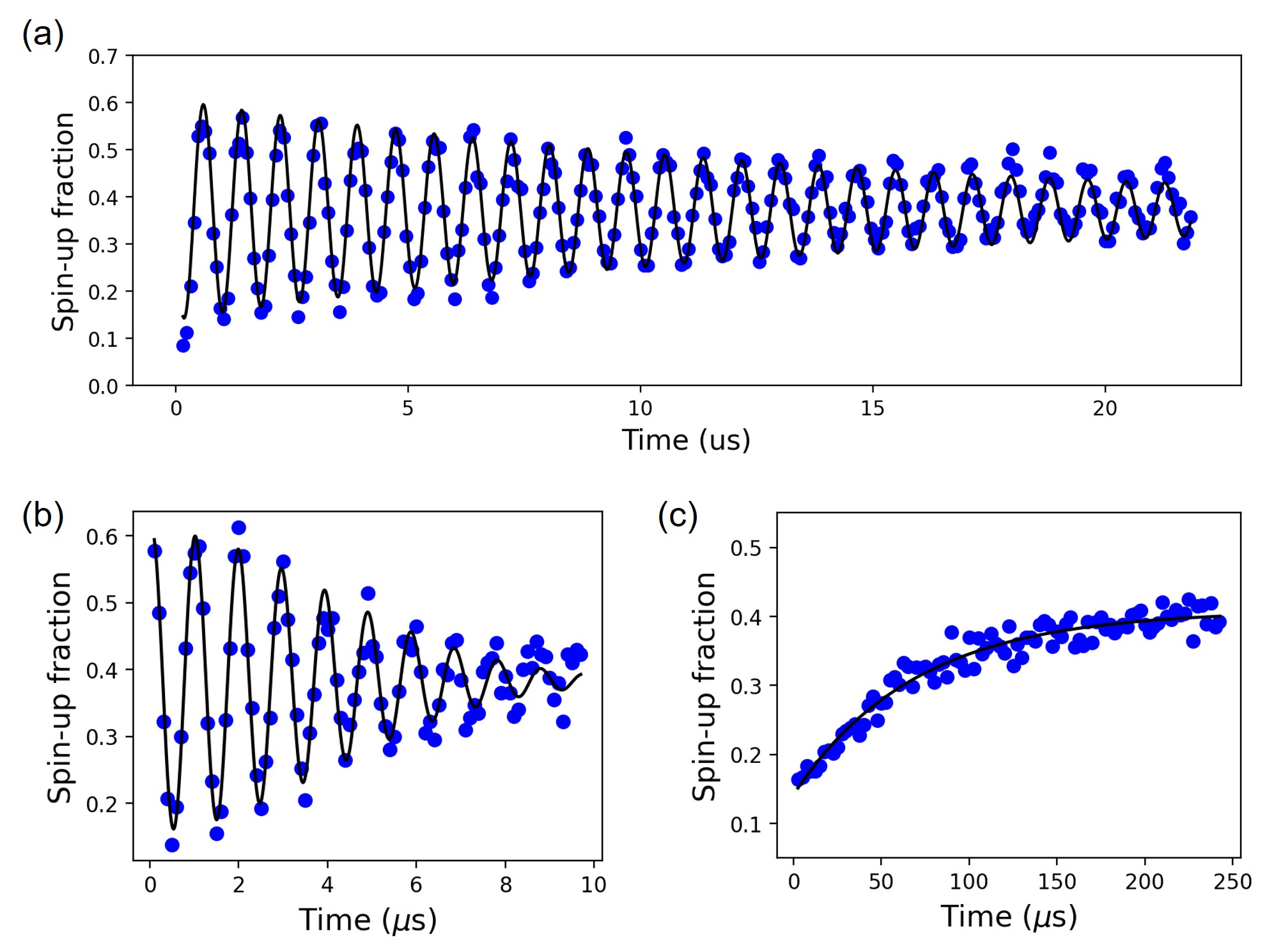}
  \caption{Bare qubit properties in the left quantum dot (Q1) under an applied magnetic field $B=0.7$ T. The qubit resonance frequency is 19.648 GHz. (a) Rabi oscillations.
  (b) Ramsey experiment.
  (c) Hahn echo experiment.
  Extracted coherence times are $T_2^{\rm{Rabi}}$ = 15.3~$\mathrm{\upmu s}$, $T_2^* = 5.5$~$\mathrm{\upmu s}$ and $T_2^{\rm{Hahn}} = 69.1$~$\mathrm{\upmu s}$.}
  \label{S1}
\end{figure}

\textcolor{black}{Below, we characterize the coherence properties of the qubit formed in the left quantum dot under an applied magnetic field of $B=0.7$ T.
Figure~\ref{S1}(a) shows the Rabi oscillations of the spin-up probability as a function of the MW pulse duration, the Rabi decay time $T_2^{\rm{Rabi}}$ = 15.3~$\mu$s.
Figure~\ref{S1}(b) presents the Ramsey experiment with an inhomogeneous dephasing time of $T_2^* = 5.5$~$\mathrm{\upmu s}$.
We perform a Hahn echo experiment, as shown in fig.~\ref{S1}(c).
The Hahn echo signal is fitted with an exponential decay, $P_{\uparrow}(t) = A\bigl[1- \exp(-t/T_{2}^{\mathrm{Hahn}})\bigr] + B$, which extends the coherence time to $T_2^{\rm{Hahn}} = 69.1$~$\mathrm{\upmu s}$.
The significant extension from $T_{2}^{*}$ to $T_{2}^{\mathrm{Hahn}}$ indicates that low-frequency noise is a dominant contribution to the qubit dephasing in this device.
The Hahn echo coherence time $T_{2}^{\mathrm{Hahn}} = 69.1$ $\mathrm{\upmu s}$ is shorter than values reported for state-of-the-art $^{28}\mathrm{Si}$ spin qubits~\cite{veldhorst2014addressable,zwerver2022qubits,steinacker2025industry}, where $T_{2}^{\mathrm{Hahn}}$ can reach several hundred microseconds to milliseconds.
This suggests that higher-frequency noise components also contribute to the decoherence in the present device.}

\section*{APPENDIX B: Power Consumption for CCD Scheme  s}

\begin{figure}[htb]
  \includegraphics[width=70mm]{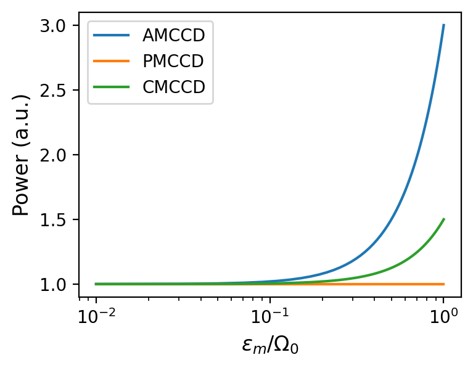}
  \caption{Comparison of power consumption among different CCD schemes.
  }
  \label{S2}
\end{figure}

Figure~\ref{S2} shows the estimated MW power required for the different CCD schemes.
For AMCCD and CMCCD, which employ amplitude modulation, the required power increases with the modulation strength $\epsilon_{\rm m}$.
AMCCD exhibits a rapid increase in power at large modulation strengths.
In contrast, PMCCD uses phase modulation, and therefore the required MW power remains constant as the modulation strength is varied.

\section*{APPENDIX C: CCD Gate infidelities}

\begin{figure}[htb]
  \includegraphics[width=\columnwidth]{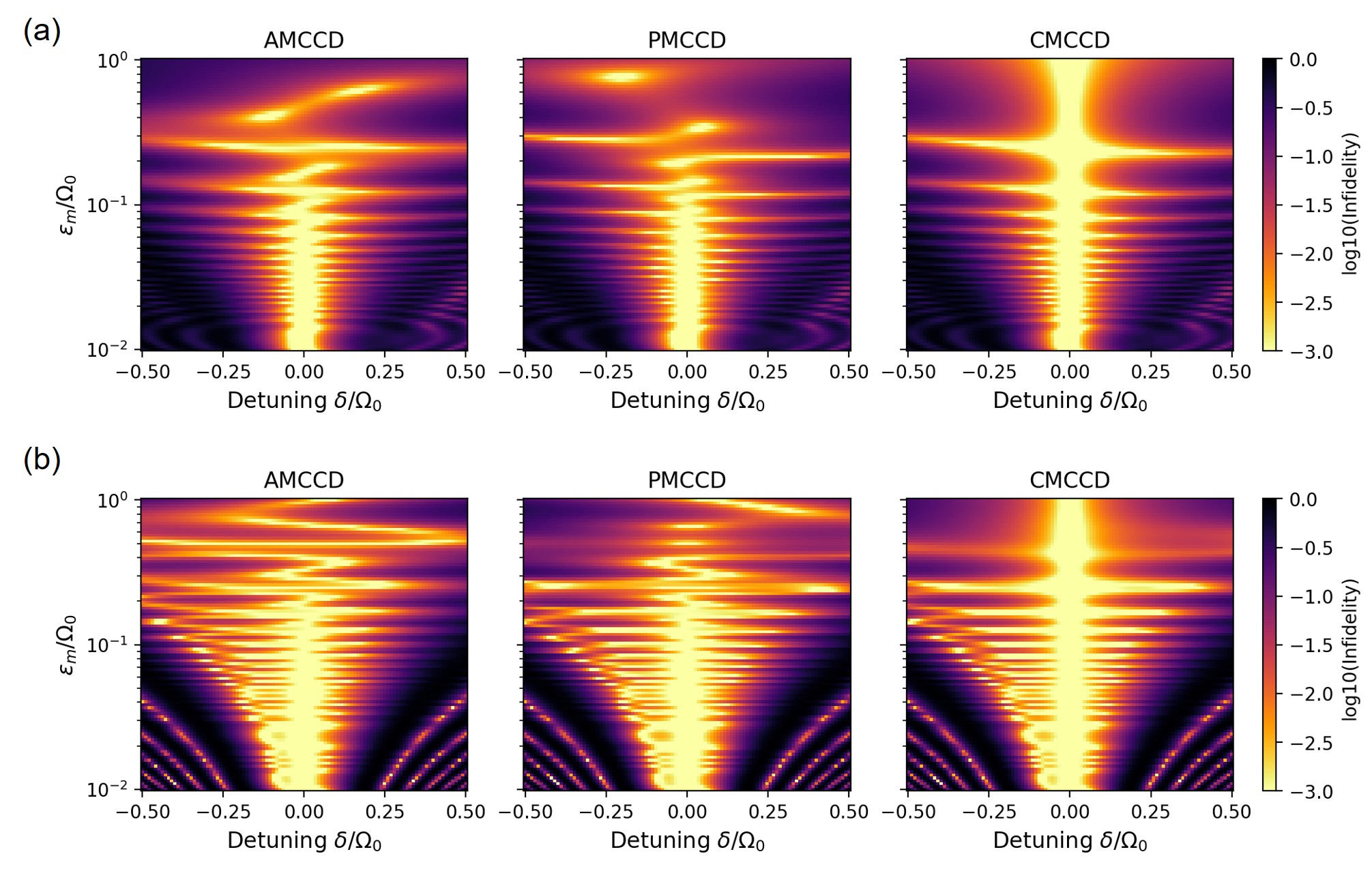}
  \caption{Gate infidelity as a function of detuning $\delta/\Omega_0$ and modulation strength $\epsilon_{\rm{m}}/\Omega_0$ for three CCD schemes: AMCCD, PMCCD, and CMCCD.
    (a) $Y_{\pi/2}$ gate and (b) $Y_{\pi}$ gate.
  }
  \label{S3}
\end{figure}

\begin{figure}[htb]
  \includegraphics[width=\columnwidth]{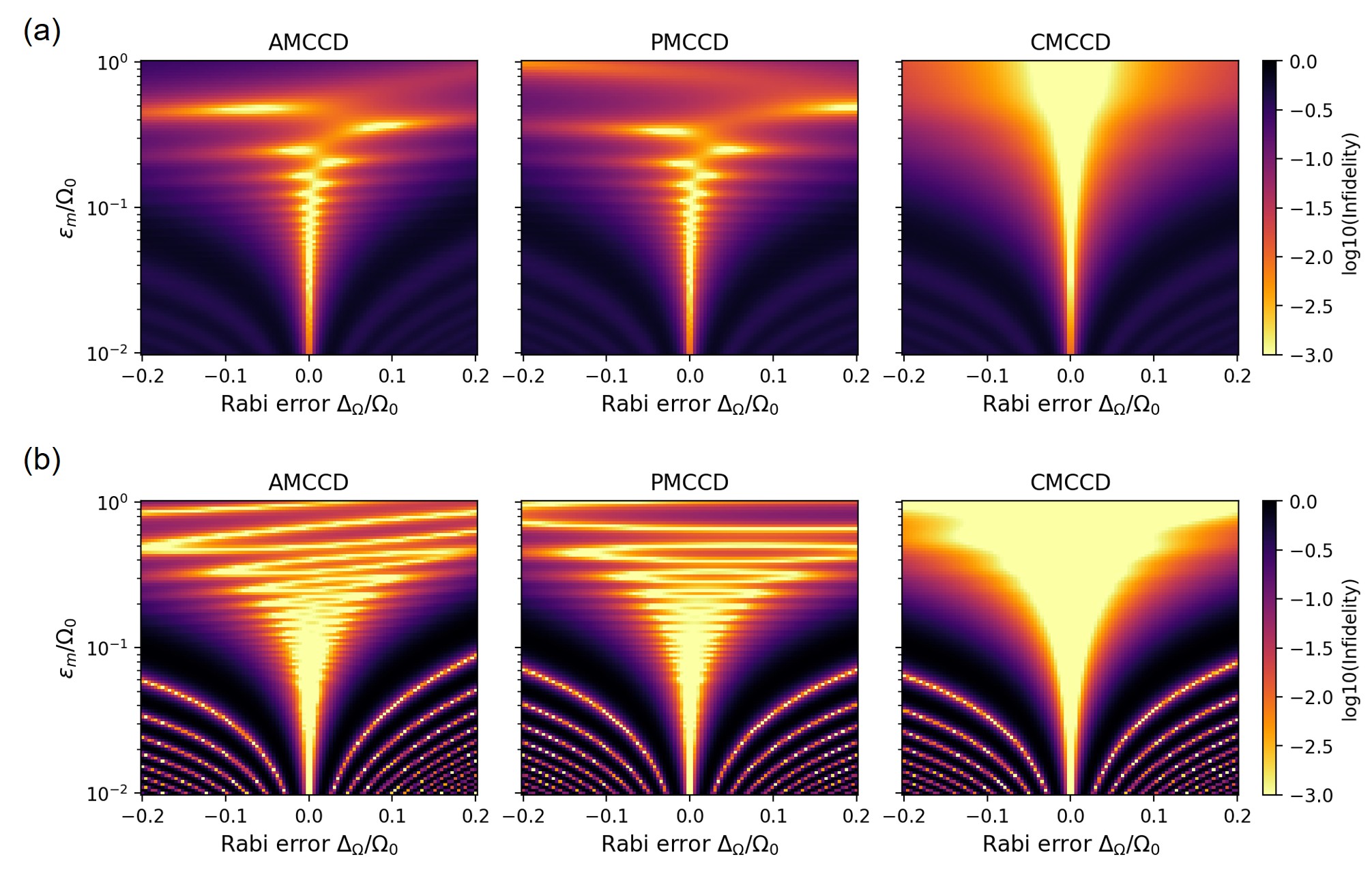}
  \caption{Gate infidelity as a function of Rabi error $\Delta_{\Omega}/\Omega_0$ and modulation strength $\epsilon_{\rm{m}}/\Omega_0$ for three CCD schemes: AMCCD, PMCCD, and CMCCD.
    (a) $Y_{\pi/2}$ gate and (b) $Y_{\pi}$ gate.
  }
  \label{S4}
\end{figure}

To evaluate the robustness of the CCD schemes, we calculate the gate infidelity as a function of the modulation strength.
Figure~\ref{S3} shows the gate infidelity as a function of the detuning $\delta/\Omega_0$ and the modulation strength $\epsilon_{\rm{m}}/\Omega_0$ for three CCD schemes: AMCCD, PMCCD, and CMCCD. In AMCCD and PMCCD, the influence of the counterrotating term becomes more significant as the modulation strength increases, which leads to a reduction in gate fidelity. In contrast, in CMCCD, the counterrotating term is eliminated, allowing high gate fidelity to be maintained over a wide range of detuning and modulation strengths.
Figure~\ref{S4} shows the gate infidelity as a function of the Rabi error $\Delta\Omega/\Omega_0$ and the modulation strength.
A similar trend is observed: AMCCD and PMCCD become increasingly sensitive to Rabi errors and modulation strengths, whereas CMCCD maintains a wide high-fidelity region. These results demonstrate that CMCCD provides improved robustness against both detuning and Rabi errors compared with conventional CCD implementations.
This robustness originates from the cancellation of the counterrotating term in CMCCD, which suppresses systematic errors associated with strong driving.

\clearpage

\bibliography{refs}

\end{document}